# Study of the relativistic charged particle beam propagation in Earth's magnetic field

Meihua Fang[1], Zheng liang[1], Yingkui Gong[2*], Jianfei Chen[1], Guiping Zhu[1], Ting Liu[2], Yu tian[2], Yu Zhou[2]

1. Nanjing University of Aeronautics and Astronautics, Nanjing, China,

2. Aerospace Information Research Institute, Chinese Academy of Sciences, Beijing, China

**Abstract:**

Relativistic charged particle beam can be used as destructive beam weapons in space for debris removal tasks. The trajectories of charged particles are affected by both electric and magnetic forces in the Earth's magnetic field. In this paper, we firstly analyzed the correlation parameters of the charged particle beam as a weapon when it propagated in the geomagnetic field. Then the models were constructed based on COMSOL Multiphysics and the IGRF model was adopted in the simulation. The gyro-radius and the related uncertainty were analyzed by simulation of the charged particle transport in the geomagnetic field at different altitudes. The charged beam spot radius divergency was also simulated. The magnetic field pinch effect can be found and can limit the beam spreading.

## 1. Introduction

The particle beam weapons, including charged particle and neutron particle beam weapons, are considered as an ideal new concept weapon equipment[1]. The charged particle beam weapons accelerate the charged particles to near-light speed and direct the beam to the target. The beam carries potentially destructive amounts of energy that can be used in space debris removal and tracking[2,3], space active detection[4,5] and space protection[6], et al.

The US Department of Defense recently wanted to test a space-based particle beam weapon in 2023[7]. And since the middle of the 20th century, the United States and the Soviet Union carried out a lot of research on the related technologies of particle beam weapons[8]. Studies on the high-energy particle transport in the magnetic field and the atmosphere were carried out. The propagation dynamics of electron beams from satellites have been studied by using analytical models and single particle tracking programs[9]. A model of Earth's atmosphere was constructed with a three-dimensional geomagnetic field based on an international geomagnetic reference field model[9]. The Monte Carlo method was used to simulate the transmission of a relativistic (1-10Mev) electron beam at the height of 200-300km. The relationship between the beam spot and beam parameters were studied[10]. Huang et al. modelled the motion characteristics of charged particles in the near-earth space and the influence of electric and magnetic fields in the magnetosphere on the charged particles drifting with different energies was analyzed[11]. Niu et al. simulated the behavior of charged particles injected into the radiation belt under different conditions[12]. Zhong et al. quantitatively analyzed the motion process of high-energy charged particles in the near-earth space[13].

However, there are still a series of unsolved problems in the charged particle beam propagation in the magnetic field as a weapon. Specifically, the divergence of the beam has great limitations on the energy and intensity of the beam transmission over long distances. It is the basic obstacle for the application of the particle beam in space[6]. And the geomagnetic field also affects the accuracy of the propagation path of the charged particles. This will also limit the weapons in outer space application[7]. This paper firstly discussed the characteristics of the charged particle propagation in the geomagnetic field. Then modelling the propagation process by using the modules in COMSOL Multiphysics was carried out. The charged particle gyration characteristics including gyro-radius and its uncertainty were discussed, and the particle beam spot radius was analyzed.

## 2. charged particle beam propagation

The propagation of the charged particle beam in outer space is affected by both the Lorentz and Coulomb forces[14] in the Earth's magnetic field. The charged particle gyration around the magnetic field line. The gyration path is decided by the geomagnetic field and affected by its disturbance. The gyro-radius, R, of high-speed particles in a uniform magnetic field is expressed as[15]:

$$R = \frac{m_0 c \beta \gamma}{qB} \tag{1}$$

where: $\gamma = (1 - v^2/c^2)^{(-1/2)}$, $\beta = v/c$, $v$ is the particle velocity, $m_0$ is the rest mass of the particle, $c$ is the speed of light, $q$ is the particle charge, and $B$ is the magnetic field strength. We can see from the equation that the smaller the magnetic field intensity, the smaller the charge-mass ratio and the larger the particle speed, the larger the radius of curvature of the path. When the particles propagate in the geomagnetic field, the Lorentz force does not do any work to the charged particles but changes the direction. The magnetic field strength in the geomagnetic field changes with the location in the earth's orbit, the same is true for gyro-radius.

The particles with the same kind of electric charge in the beam repelled with each other according to the Coulomb's raw which causes the beam divergence. Divergence can be simply expressed by the ratio between the beam's rate of expansion and rate of travel. Beam divergence is a great limitation for long-distance beam transmission, and is the basic obstacle for the space application of charged particle beam[16]. According to relativistic particle dynamics, the beam spot radius a at the distance from the injection source z is determined by equation (1)[16] as follows:

$$aF(ln(a/a_0)^{1/2}) \; ; \; (z/\beta\gamma)(I/I_A)^{1/2} \qquad (2)$$

where $a_0$ is the initial beam spot radius, F is the Dawson integral, $I$ is the beam current, $I_A$ is the Alfven current. It can be seen from the formula that the radius of the beam spot is proportional to the distance z and current intensity.

In addition, the magnetic field has a "focusing" effect which can limit the spreading of the charged particle beam. Charged particles in a solenoidal magnetic field experience an axial Lorentz force that causes them to move in helical paths along the solenoid axis. This motion effectively constrains the radial movement of the particles, providing a focusing effect[17]. The radial force on an electron at a radial distance r from the solenoid axis is expressed as:

$$F_r = 2\frac{mc^2}{\beta\gamma^2}\frac{I}{I_A}\frac{r}{R^2}$$

Where $F_r$ is the radial force and this force could oppose the Lorentz force due to the solenoid field. due to the solenoid field In the Earth's magnetic field, charged particles move in helical paths along the magnetic field lines, which constrains their motion perpendicular to the magnetic field and helps prevent the particle beam from dispersing. Although in space, the Earth's magnetic field is relatively weak and its focusing effect is very limited, it should be still be considered.

### 3. Modelling and analysis

#### 3.1 Charged particle gyration

In this section, we used COMSOL Multiphysics software to simulate the process of charged particle propagation in the geomagnetic field. International Geomagnetic Reference Field (IGRF) was adopted in the simulation, as shown in Figure 1. The particle beam injection location and the corresponding geomagnetic field strength are shown in Table 1. $R_e$ is the earth radius listed in the table. The pitch angle of the particle with the Earth's magnetic field is 90 degrees.

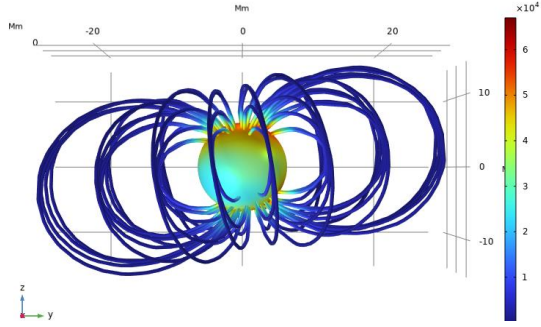

Figure 1 IGRF model in the simulation.

Table 1 Simulate parameters

| Location name | Location | Magnetic field strength /T |
| --- | --- | --- |
| LEO-P | $(1.16, 0, 0) R_e$ | $1.881 \times 10^{-5}$ |
| MEO-P | $(4.14, 0, 0) R_e$ | $4.021 \times 10^{-7}$ |
| GEO-P | $(6.65, 0, 0) R_e$ | $9.881 \times 10^{-8}$ |

10 MeV proton beam propagation in the geomagnetic field was carried out firstly and the trajectories were shown in Figure 2. It is obviously that the gyro-radius of LEO-P < MEO-P < GEO-P in the simulation. The resulted gyro-radius was shown in Figure 3. As it can be seen from the figure, the gyro-radius of 10 MeV proton in the geomagnetic field is periodically. The gyro-radius period, $T_c$, can be expressed as:

$$T_c = \frac{2\pi m_0}{q|\boldsymbol{B}|} \quad (3)$$

The equation shows that gyro-radius period is the reciprocal of the geomagnetic field strength. The gyro-radius period in the position of GEO-p ($T_c = 0.76s$) is more than two orders of the magnitude higher than the gyro-radius in the position of LEO-P ($T_c = 0.0038s$)。 And the gyro-radius and its variant range in in GEO-P is larger than MEO-P and LEO-P. The gyro-radius values and period for 10 MeV proton and electron are shown in Table 2 and Table 3 respectively. Due to the small mass of the electron, the gyro-radius of the electron beam is much smaller than that of the proton with the same energy. The gyro-radius is small compared to the spatial scale for satellite orbit especially for 10 MeV electron beam. The magnetic field resulting from the earth is not uniform. A charged particle gyrates in geomagnetic field. When it moves to a point with a stronger magnetic field strength, the gyro-radius becomes smaller. Conversely, when it reached a point with a weaker magnetic field, the gyro-radius becomes larger. As a result, the gy-radius for proton and electron is not uniform but vary with the geomagnetic field.

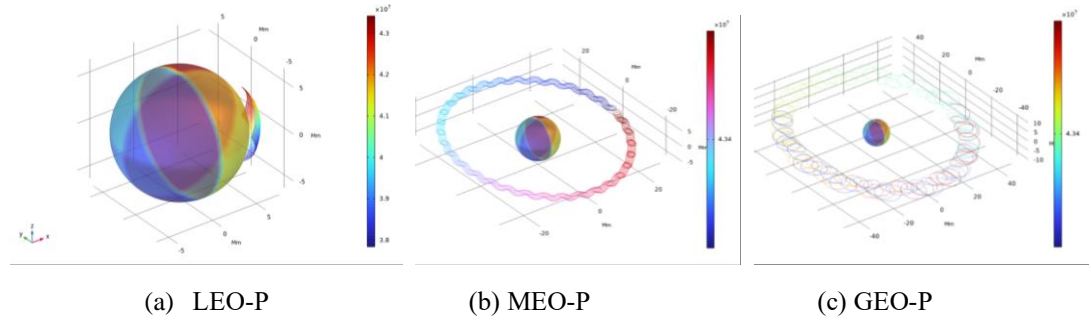

(a) LEO-P    (b) MEO-P    (c) GEO-P

Figure 2 the trajectories of 10 MeV proton propagation in the geomagnetic field

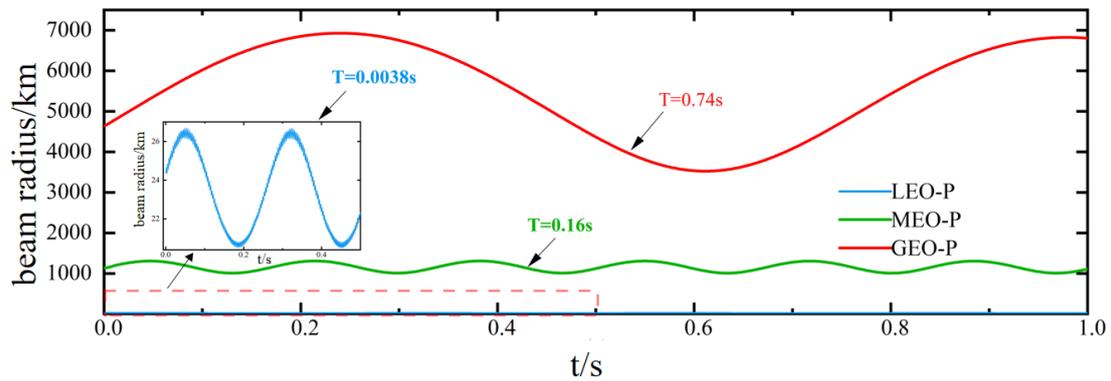

Figure 3   The gyro-radius of 10 MeV proton.

Table 2 The gyro-radius range and period of the 10 MeV proton in the first period

| Position | Maximum gyro-radius /km | Minimum gyro-radius /km | Gyro-radius period /s |
|---|---|---|---|
| LEO-P | 26.698 | 26.159 | 0.0038 |
| MEO-P | 1308.207 | 1013.102 | 0.1600 |
| GEO-P | 6927.942 | 3525.882 | 0.7400 |

Table 3 The gyro-radius range and period of 10 MeV electron in the first period

| Position | Maximum gyro-radius /km | Minimum gyro-radius /km | Cyclotron period /s |
|---|---|---|---|
| LEO-P | 1.380 | 1.378 | 1.44 x 10 |
| MEO-P | 87.948 | 86.248 | 1.85 x 10 |
| GEO-P | 363.268 | 345.894 | 7.44 x 10 |

The Earth's magnetic field is not static but constantly disturbing. The geomagnetic field is disturbed by interplanetary magnetic field and solar activity. This will cause changes in particle trajectory and result in uncertainty for the particle gyration path. The relationship between the uncertainty of beam propagation and the geomagnetic field is calculated by[16]:

$$\Delta R = R(\Delta B / B) \qquad (4)$$

where: $\Delta R$ is the uncertainty at the gyro-radius R; $\Delta B$ Is the magnetic field strength change value by disturbing. $\Delta B / B$ is defined as the uncertainty of the magnetic field.

As can be seen from equation (4), the uncertainty of the gyro-radius for the particle beam increases with the increase of the gyro-radius and the uncertainty of the magnetic field. If the geomagnetic field uncertainty is estimated to be 3% and the gyro-radius is 100km, then the uncertainties of gyro-radius will be 3 km。This scale is larger than most of the space debris and satellites. In our study, the magnetic field was calculated by IGRF model, and this model is a mathematical representation for internal part of Earth's magnetic field. There is a uncertainty between IGRF model value and the detection data[18]. And for high altitude, the external magnetic field should be added[19]。 In addition, a nuclear explosion in the stratosphere would also interact with the geomagnetic field, and the effect would be large and unpredictable. Considering all these factors, the uncertainties in gyro-radius and gyration path for a charged particle could also be large and unpredictable.

### 3.2 Charged particle beam divergence

In this section, the electron beam divergence in the geomagnetic field was modelled with combined AC/DC module and particle tracking module in COMSOL Multiphysics. The two-way couplings between particles and magnetic field were used. In the simulation, the particles were subjected to Lorentz forceand and Coulomb force, with the state changes of the particles being recorded at each step. The Lorentz force can be expressed as: $F = zev \times \mathbf{B}$. The Coulomb force can be expressed as $F = \frac{e^2}{4\pi\epsilon_0}\sum_{j=1}^{N} zz_j \frac{r-r_j}{|r-r_j|^3}$.where $\epsilon_0$ is permittivity of free space, $N$ is the electron numbers in the beam.

The magnetic field intensity calculated by IGRF model was set to a geographical location (Beijing, China, Height 400 km). The geometry model constructed in the simulation was shown in Figure 4. The electron beam is emitted from the cylindrical surface and travels along the Z direction. The beam radius is 0.01 meters, and the beam current is 3 A.

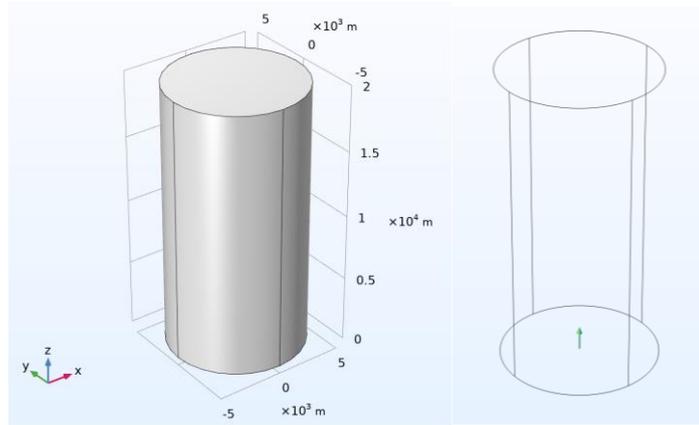

Figure 4 The simulation geometry.

The trajectory of the particle in the magnetic field was shown in Figure 5. It could be observed from the figure that the particle moved in a helical path around the magnetic field lines. The beam radius size with and without geomagnetic field over time was shown in Figure 6. It was obviously from the figure that the geomagnetic field inhibited the diffusion of the electron beam. Additionally, when the beam moved in the geomagnetic field, its' radius increased firsly, then decreased, and increased again, exhibiting periodic changes. These periodic changes in beam radius were related to the magnetic field induced force at the position of the particle's helical motion, as shown in Figure 7. In this comsol study, cpt.max(qx) means the maximum x position of the particle beam, cpt.min(qx) means the minimum x position of the particle beam, and so on for the (qy). It was assumed that (cpt.max-cpt.min) equals the maximum diameter for the beam spot. Overall, the beam diameter change in the X direction was inversely related to the change in magnetic force in the Y direction, and the beam radius in the Y direction was positively correlated to the magnetic force in the X direction. That why the charge of the beam radius for x and y direction was different.

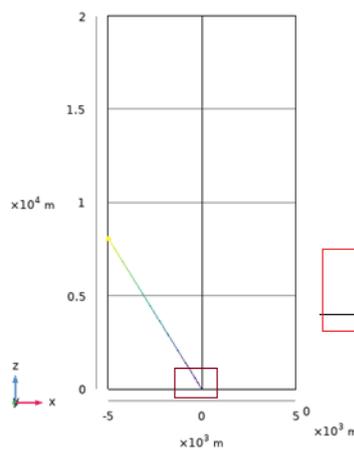

Figure 5 the trajectory of the electron beam

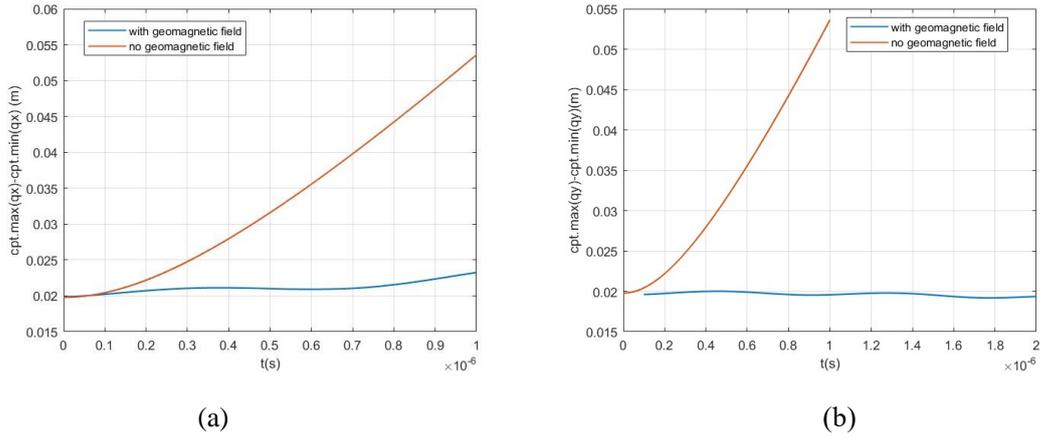

(a)          (b)

Figure 6 the beam radius of the electron beam with and without earth's magnetic field. cpt.max(qx) and cpt.max(qy) is the maximum value of x and y for the particle position in the trajectory. cpt.min(qx) and cpt.min(qy) is the minimum value of x and y for the particle position in the trajectory

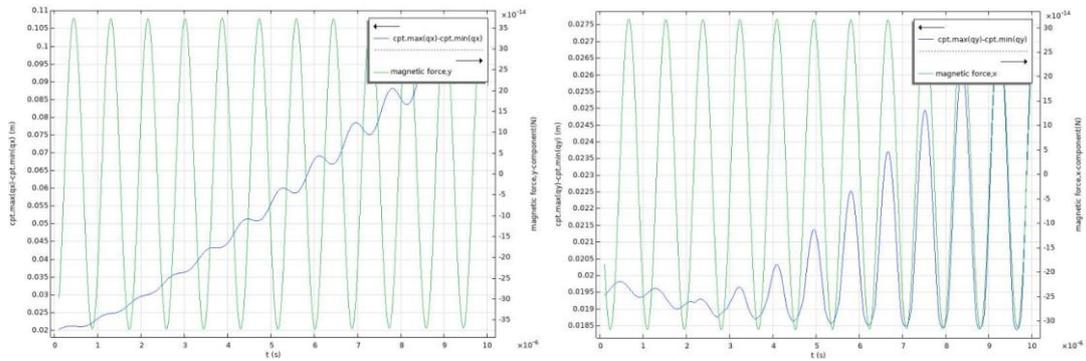

Figure 7 The max diameter along x axis and y axis

## 4. Conclusion

Particle beam weapon is a type of directed-energy weapon and remain at the research stage. Space based charged particle beam weapon have serials of unsolved problems in the process of charged particle propagating in the geomagnetic field. One problem is the charged particle trajectory path and its uncertainty, the other is the particle beam divergence.

In this work, models based on COMSOL Multiphysics was constructed to simulate the charged particle transport in the geomagnetic field. It is obviously that the gyro-radius in the geomagnetic field is periodicity. The higher the altitude, the larger the period. The electron beam have smaller gyro-radius compared to proton with the same energy. The uncertainty of gyro-radius contributed from the magnetic disturbing and will dramatically affect the particle trajectory. In addition, the geomagnetic field pinch effect exist in the simulation, however, the beam divergence is dramatically and will limit the destructive effect of the beam. The charged particle trajectory path disturbing and beam divergence for different energy particles with inner geomagnetic field and outer geomagnetic field should be modelled and studied in detail in further researches.